\documentclass[12pt]{article}
\usepackage[utf8]{inputenc}
\usepackage[english]{babel}
\usepackage{amsmath, amssymb,graphicx}
\usepackage{braket}
\usepackage{color}
\usepackage{ulem}
\usepackage{yfonts}
\usepackage{caption}
\usepackage{subcaption}
\pdfoutput=1
\usepackage{jheppub}
\newcommand{\be}{\begin{equation}}
\newcommand{\ee}{\end{equation}}
\newcommand{\bea}{\begin{eqnarray}}
\newcommand{\eea}{\end{eqnarray}}
\newcommand{\nn}{\nonumber}

\def\({\left(} \def\){\right)}
\def\[{\left[} \def\]{\right]}

\def\Re{\text{Re}}
\def\Im{\text{Im}}

\title{}

\def\Pf{\text{Pf}}

\newcommand{\fx}{\mathfrak{x}}
\newcommand{\fy}{\mathfrak{y}}

\newcommand{\fT}{\mathfrak{T}}
\newcommand{\fC}{\mathfrak{C}}
\newcommand{\fH}{\mathfrak{H}}

\newcommand{\fS}{\mathfrak{S}}

\usepackage{setspace}

\title{Notes on the SYK model in real time
}

\author{Irina Aref'eva and Igor Volovich}

\affiliation{Steklov Mathematical Institute, Russian Academy of Sciences,\\Gubkina str. 8, 119991, Moscow, Russia}

\emailAdd{arefeva@mi.ras.ru, volovich@mi.ras.ru}

\abstract{ Nonperturbative formulation of the Sachdev-Ye-Kitaev (SYK) model is discussed.
 The partition  function of the model can be represented as a functional integral
over the Grassmann variables in Euclidean time which is well defined but it diverges after the  transformation to the fermion bilocal fields.  We point out that the generating functional of the SYK model
in {\it real time} is well defined even after the  transformation to the bilocal fields and it  can be used for  nonperturbative investigations of its properties.   The  SYK model in zero dimensions is studied, its large $N$ expansion is evaluated and  phase transitions are investigated.

}

\begin{document}

\maketitle
\flushbottom
\newpage
\section{Introduction}
\label{sec:intro}
The Sachdev-Ye-Kitaev (SYK) model \cite{Sachdev,Sachdev:2015efa,Kitaev} has been considered recently as an interesting example of the solid state system which could admit a nontrivial holographic description,
\cite{Polchinski:2016xgd,JSY,MS,Maldacena:2016upp,Bagrets:2016cdf,Jensen:2016pah,Engelsoy:2016xyb,Cotler:2016fpe,Gross:2017hcz,Gross:2017aos,KitSuh}.
The ordinary approach to the study of the partition function of the SYK model is based on the  transformation to the fermion bilocal fields  and then the using of the saddle point method to get the $1/N$ expansion. However in terms of bilocal fields
one gets the functional integral that is divergent and it is difficult to use it for a nonperturbative formulation.

 We point out  that the generating  functional of the SYK model
in  the  {\it real time} is well defined even after the  transformation to bilocal fields and it  can be used for  nonperturbative investigations of its properties.

A simple analogue of the Euclidean partition function of the SYK model can be described as
the Gaussian integral with the random coupling constant $g$
\be\label{Fphi4}
\frac{1}{\sqrt{2 \pi }}\int dx dg\exp\{-\frac{g^2}{2}  \}   \exp\{\pm gx^2\}=\int dx \exp\{\frac{1}{2}x^4\},
\ee
which is divergent. The problem of giving meaning to the functional integral of the SYK model in the bilinear variables representation by a suitable choosing  of complex contours has been  discussed in \cite{Cotler:2016fpe}
and in \cite{KitSuh} with the conclusion that  still there are open questions.

In the    "real time" formulation discussed in this paper, we have
\be
\label{TRphi4}
\frac{1}{\sqrt{2 \pi }}\int dx dg\exp\{-\frac{g^2}{2}  \}   \exp\{i gx^2\}=\int dx \exp\{-\frac{1}{2}x^4\},
\ee
which  is well defined.

We propose to define
the generating  functional   in the  SYK model in zero dimensions with  $M$ replicas in "real time" as  the Grassmann (Berezin) integral over anticommuting variables

\bea
&&\,\,\,\,\,\,\,\,\,\,Z(A,J)=\int d\mu ({\bf J})\,\int \, d\chi\,e^{i{\cal A}[\chi, J]},
\\
{\cal A}[\chi, J]&=&\frac{i}2\sum_{\alpha,\beta,j}\chi_j^{\alpha}A_{\alpha\beta}
\chi_j^{\beta}+\sum_{\alpha}\sum_{i<j<k<l}\sqrt{\frac{3!J^2}{N^3}}J_{ijkl}\chi_i^{\alpha}
\chi_j^{\alpha}\chi_k^{\alpha}\chi_l^{\alpha}.\label{S}
\eea
where $\chi_j^{\alpha}$ are the Grassmann variables, $i,j,k,l=1,...N,\,\,\alpha,\beta=1,.. M$, $J>0$, $(A_{\alpha\beta})
$  is an antisymmetric matrix with real entries and   $d\mu ({\bf J})$ is a Gaussian probability measure and ${\bf J}=(J_{ijkl})$ are  random variables with
zero mean $<J_{ijkl}>=0$, and variance  is\\ given by  $<J_{ijkl}^2>=3!J^2/N^3$. 

Performing the integration over $d\mu ({\bf J})$ one gets
\bea
Z(A,J)&=&\int d\chi\exp
 \left\{-\frac12\sum_{\alpha,\beta,j}\chi_j^{\alpha}A_{\alpha\beta}
\chi_j^{\beta}-  \frac{NJ^2}{8}\sum_{\alpha,\beta}\Big( \frac{1}{N}\sum_j\chi_j^{\alpha}
\chi_j^{\beta}\Big)^4   \right\}.
\label{Z0dim}
\eea
Note that $Z(A,J)$ is a polynomial at $J^2$ and $A_{\alpha\beta}$ because there is only a finite number
of the Grassmann variables.
A similar natural definition can be done also for the 1-dim SYK model, which can be called the SYK model in  real time, see below.

The expression \eqref{Z0dim} can be written also by using the bilocal variables
\bea
Z(A,J)=\left(\frac{N}{2\pi}\right)^{\frac{M(M-1)}{2}}\int d\Sigma\, dG\,\left( \mathrm{Pf}(A+i\Sigma)\right)^N\,\exp\left\{N\sum_{\alpha<\beta}
\left[-\frac{J^2}{4}  G_{\alpha\beta}^4+i\Sigma_{\alpha\beta}G_{\alpha\beta}\right]\right\}.\nn\\
\label{Z0M}\eea
Here Pf is the Pfaffian and $ (G_{\alpha\beta})$ and $ (\Sigma_{\alpha\beta})$ are antisymmetric matrices with real entries.
Note the presence of the factor $\exp\{-J^2  G_{\alpha\beta}^4/4\}$ which provides the convergence of the integral.

We will study the large $N$ behaviour of the particular case of the 0-dim SYK model with two replicas (M=2)
which is just the  integral
\be
Z_q=\frac{N}{2\pi}\int dy (A+iy)^N\int dx \exp \{N(-\frac{J^2}{q}x^q +ixy) \}\label{GF}
\ee
with $q=2$ and $q=4$. Here $x,y$ are real variables and  $A$ is a real constant, $J>0$ and $N$ is a natural number.
The similar  integral as a toy model for a suitable choice of the contour in the SYK model has been  considered in \cite{Cotler:2016fpe}. 
In the case $q=2$ the expression \eqref{GF} after rescaling of $x$ and $y$ coincides  up to a constant  with the Hermite polynomial $H_N\left(\sqrt{2N}\frac{A}{2J}\right)$. Using the asymptotic behavior of the Hermite polynomials
for $N\to \infty$ we observe the phase transition at  $A/2J=\pm 1$ in this model. Similarly we find the phase transition for $q=4$.

The generating functional  of the 1-dim SYK model with  $M$ replicas can be represented in the form of the Grassmann  integral over anticommuting variables
$\chi_j^{\alpha}=\chi_j^{\alpha}(\tau),\,\tau$ is on real line, as

\be
<Z^M>=\int d\mu ({\bf J})\int {\cal D}{\bf \chi}e^{i{\cal A}},
\ee
where the action is
\be
{\cal A}=\sum_{\alpha}\int d\tau(\frac{i}{2}\sum_j\chi_j^{\alpha}
\partial_{\tau}\chi_j^{\alpha}
+\sum_{i<j<k<l}\sqrt{\frac{3!J^2}{N^3}}J_{ijkl}\chi_i^{\alpha}
\chi_j^{\alpha}\chi_k^{\alpha}\chi_l^{\alpha}).
\ee
One has
$$<Z^M>=\int {\cal D}{\bf \chi}\exp(-\frac{1}{2}\sum_{\alpha,j}\int d\tau\chi_j^{\alpha}
\partial_{\tau}\chi_j^{\alpha}-
\frac{NJ^2}{8}\sum_{\alpha,\beta}\int d\tau
\int d\tau^{\prime}(\frac{1}{N}\sum_j
\chi_j^{\alpha}(\tau)
\chi_j^{\beta}(\tau^{\prime}))^4).
$$

The generating functional can be also written in terms of the bilocal variables

\be\label{IM}
<Z^M>=\int  {\cal D }\Sigma\,{\cal D}G\exp(NI),
\ee
where
\bea
&&\,\,\,\,\,\,\,\,\,\,\,\,\,\,\,\,\,\,\,\,\,\,\,\,\,\,\,\,
\,\,\,\,\,\,\,\,\,\,\,\,\,\,\,\,I=\ln  \mathrm{Pf}(\partial_{\tau}+i\Sigma)\\\nn&&+\sum_{\alpha <\beta}\int d\tau
\int d\tau^{\prime}\Big(i\Sigma_{\alpha\beta}(\tau,\tau^{\prime})
G_{\alpha\beta}(\tau,\tau^{\prime})-\frac{J^2}{4}G_{\alpha\beta}(\tau,\tau^{\prime})^4\Big).
\eea
Here $G=(G_{\alpha\beta}(\tau,\tau^{\prime})),\,\Sigma=(\Sigma_{\alpha\beta}
(\tau,\tau^{\prime}))$ and $G_{\alpha\beta}(\tau,\tau^{\prime}),\,\Sigma_{\alpha\beta}
(\tau,\tau^{\prime})$ are antisymmetric real valued functions (bilocal fields).
    Note that the factor $\exp\{-NJ^2G_{\alpha\beta}(\tau,\tau^{\prime})^4/4\}$ provides the convergence of the integral.

The paper is organized as follows.
In Section \ref{sec:0dim} we obtain the representation  \eqref{Z0M}.
In Section 3 we consider the simplest case of  the $q=2$ model and using the relation 
of the model with the Hermite polynomials  investigate the behaviour of the model in the large $N$ limit. In Section 4
using the steepest  descent method we investigate the behaviour of the 
$q=4$ model in the large $N$ limit. We conclude in Section 5 with the discussion.
$$\,$$

\section{The 0-dim SYK model}\label{sec:0dim}

The generating functional for 0-dim SYK model is given by the formula \eqref{Z0dim}.
Let us demonstrate that it can be represented in the form \eqref{Z0M}.
To this end we use the Fourier transform. If $f(x)$ is an integrable fast decreasing function on the real axis then
the following identity for the inverse Fourier transform holds
\begin{equation}
\label{fFT}
f(\Theta)=\frac{1}{2\pi}
\int_{-\infty}^{\infty}dy\int_{-\infty}^{\infty}dx\,f(x)e^{ iy(x-\Theta)}.
\end{equation}
 This formula is valued not only in the case when $\Theta$ is the real variable, but also in the case when $\Theta $ is a nilpotent element of the Grassmann algebra, in particular,
 for
 $\Theta=\Theta_{\alpha\beta}$,
 \be\Theta_{\alpha\beta}=\frac{1}{N}\sum_j\chi_j^{\alpha}
\chi_j^{\beta}.
\ee
It is assumed that the functions $f(\Theta)$ and $e^{-iy\Theta}$  are  understood as power series in $\Theta$.

Note that we don't use such expressions as $\delta (x-\Theta)$. In principle such an object can be defined by using superanalysis on the Banach algebras developed in
\cite{Vlad-Vol-1} but at this point we don't need it.
\subsection{Bilocal variables in the 0-dim model}
We start from the generating functional in the real time formulation given by representation \eqref{Z0dim}. Using the representation \eqref{fFT} for $f( \Theta_{\alpha\beta})=\exp\{-NJ^2  \Theta_{\alpha\beta}^4/4\}$,
where $\Theta_{\alpha\beta}= \frac{1}{N}\sum_j\chi_j^{\alpha}
\chi_j^{\beta}$, we get
\bea\nn
e^{-\frac{NJ^2}{4}  \Theta_{\alpha\beta}^4}=\int \prod_{\alpha<\beta}\frac{d\Sigma_{\alpha\beta}}{2\pi}\int \prod_{\alpha<\beta}dG_{\alpha\beta}
\exp\{-\frac{NJ^2}{4}  G_{\alpha\beta}^4 + i\Sigma_{\alpha\beta} (G_{\alpha\beta}-\Theta_{\alpha\beta}) \}\eea
and
\bea
Z(A,J)&=&\int \prod_{j,\alpha} d\chi^\alpha_j \int \prod_{\alpha<\beta}\frac{Nd\Sigma_{\alpha\beta}}{2\pi}\int \prod
_{\alpha<\beta}dG_{\alpha\beta}
\exp\left\{N\sum_{\alpha<\beta}\Big[-\frac{J^2}{4}  G_{\alpha\beta}^4 + i\Sigma_{\alpha\beta}G_{\alpha\beta}\Big]\right\}\nn\\
&\cdot&\exp \left\{-\frac12\sum_{\alpha,\beta,j}
A_{\alpha\beta}\chi_j^{\alpha}\chi_j^{\beta} -
 \frac{i}{2}\sum_{\alpha,\beta,j}\Sigma_{\alpha\beta}\chi_j^\alpha\chi_j^\beta
 \right\}.
\label{Z0dimr}
\eea
Integrating over the Grassmann variables we get
\bea
Z(A,J)&=& \int \prod_{\alpha<\beta}\frac{N\,d\Sigma_{\alpha\beta}}{2\pi}\int \prod
_{\alpha<\beta}dG_{\alpha\beta} \,\left(\Pf\left(i\Sigma_{\alpha\beta}+A_{\alpha\beta}\right)\right)^{N}
\nn\\  &\cdot &\exp\left\{N\sum_{\alpha<\beta}\Big[-\frac{J^2}{4}  G_{\alpha\beta}^4+i\Sigma_{\alpha\beta}G_{\alpha\beta}\Big]\right\}.\label{Z0dimrrr}
\eea

\section{Zero-dim SYK model with 2 replicas}
\subsection{Real time for zero-dim SYK model with 2 replicas}
Let us consider the 0-dim SYK model with 2 replicas. In this case we have only the following  variables $G_{12}= x$, $\Sigma_{12}=y$ and $A_{12}= A$ and the generating functional has the form
\be
Z_q(A,J)=\frac{N}{2\pi}\int dy\int dx\, (iy+A)^N
\exp\{N(-\frac{J^2}{q}x^q+ixy)\}, \,\,\,\,\,\,\,q\geq 2.
\ee
We will study its asymptotic behaviour as $N\to \infty$.
The case  $A=0$ corresponds to the vacuum functional, $A\neq 0$  stands for the generating functional.

  \subsection{Quadratic model and Hermite polynomials}
  Let us now  consider the generating  functional for the quadratic case
\be\label{FG2c}
Z_2(A,J)=\frac{N}{2\pi}\int dy\int dx\, (iy+A)^N
\exp\{N(-\frac{J^2}{2}x^2+ixy)\},\,\,\,\,\,J>0.
\ee
 The following formula  holds
\bea\label{FG2N}
Z_2(A,J)&=&2\left(\frac{J}{\sqrt{2N}
  }\right)^{N} H_N(\sqrt {2N}\frac{A}{2 J}),\label{f2Her}
\eea
where $H_N(x)$ are the Hermite polynomials, $N=1,2,...$. Indeed,
\bea\label{FG2N}
Z_2(A,J)&=&\frac{1}{J}\sqrt{\frac{N}{2\pi}}\int dy\, (iy+A)^N e^{-\frac{N y^2}{2
   J^2}}\label{f2IR}
\eea
and by using the representation
  \begin{eqnarray}\label{HIR}
H_N(x) &=& \frac{2^N}{\sqrt{\pi}} \int_{-\infty}^\infty dt\, (x+it)^N \,e^{-t^2}
\end{eqnarray}
after the change of variables  $y=
   Jt\sqrt{\frac{2} {N}}$ in \eqref{f2IR}
we obtain \eqref{f2Her}.

 \subsection{Phase transition}

  It was found by Plancherel and Rotach that there are two regions on the half-line $x>0$ where the Hermite polynomials $H_N((2N+1)^{\frac{1}{2}}x)$ have different asymptotic behaviour as $N\to\infty$ for $0<x<1$
  and for $1<x<\infty $, see \cite{Szego}.  Recall that the Hermite polynomials satisfy the symmetry condition $H_N(-x)=(-1)^N H_N(x)$.
  In our case we have to use a slightly modified asymptotic, that has been found later in \cite{Wyman} and more recent paper \cite{Dominici}. In this case also there is different
  asymptotical behavior of the Hermite polynomials $H_N(\sqrt{ 2N}\cdot x)$ for $|x|>1$ and $|x|<1$.  
  The critical points for the Hermite polynomials $H_N(\sqrt{ 2N}\cdot \frac{A}{2 J})$ are $A/2J=\pm 1$.   We have:
  \begin{itemize}
   \item For $A/2 J>1$, using parametrization
   \bea
 \frac{A}{2 J}= \cosh \Psi,\eea
 we have for $\psi>0$ asymptotically for $N\to \infty$
   \bea
 Z_2(A,J)&\sim & Pe^{NF}\nn\\&\equiv & 2
   \exp\left[  \frac{N}{2}\left(e^{-2 \Psi }+2\Psi+2\ln J\right)+\frac12\Psi
  \right] \sqrt{\frac{1}{2\sinh \Psi}  },
  \label{UHLf}
\eea
where 
\be
F= \frac{1}{2}\left(e^{-2 \Psi }+2\Psi+2\ln J\right).
\ee
Note that $F$ could take positive and also negative values. 

For $\psi <0$ with the same parametrization we have
  \bea
 Z_2(A,J)&\sim &2
   \exp\left[  \frac{N}{2}\left(e^{2 \Psi }-2\Psi+2\ln J\right)-\frac12\Psi
  \right] \sqrt{\frac{-1}{2\sinh \Psi}  }.
\label{UHLf}
\eea
 \item  For $0<A/2 J<1$, using parametrization
   \bea
 \frac{A}{2 J}= \cos \psi,\eea
 we have for $\psi>0$
   \bea
 Z_2(A,J)&\sim & Pe^{NF}\equiv 2
\exp\left\{  \frac{N}{2}\left[  2\ln J  +\cos 2\psi \right]  \right\} \nn\\
&\times&\cos\left\{  N\left[  \frac{1}{2}(\sin 2\psi
-2\psi)\right]  +\frac{\pi}{4}-\frac{\psi}{2}\right\}\sqrt{\frac
{2}{\sin \psi  }},  \label{UOH}
\eea
where
\be
F= \frac{1}{2}(  2\ln J  +\cos 2\psi) .
\ee
For $\psi <0$ with the same parametrization we have  the similar asymptotics with
$\psi\to -\psi
$ in \eqref{UOH}, i.e.
 \bea
 Z_2(A,J)&\sim &2
\exp\left\{  \frac{N}{2}\left[  2\ln\left(
J\right)  +\cos\left( 2\psi)\right)  \right]  \right\} \nn\\
&\times&\cos\left\{  N\left[  \frac{1}{2}(\sin 2\psi
-2\psi)\right]  -\frac{\pi}{4}+\frac{\psi}{2}\right\}\sqrt{-\frac
{2}{\sin \psi  }}.  \label{UOHm}
\eea
\item
 For  $A\approx 2J$  one uses the parametrizaton
$\frac{A}{2 J}=1-\frac{z}{2^{\frac {1}{2}}3^{\frac {1}{3}}N^{\frac {1}{6}}\sqrt {2N+1}}$
 with $z$ complex and   bounded, one has
 \bea
 Z_2(A,J)&\sim &  (2J\sqrt{1/e})^N (2\pi )^{\frac12}\,N^{\frac {1}{6}}e^{\frac {x^2}{2}}
   \left(\operatorname {Ai} \left(-3^{-{\frac {1}{3}}}z\right)+O\left(N^{-{\frac {2}{3}}}\right)\right).
 \label{TRR}
 \eea
 
 \end{itemize}
Note that the same asymptotics \eqref{UHLf} and \eqref{UOH} can be found by  using the steepest  descent  method.

The half-plane $(A,\,\,J>0)$ is divided by the curves $F(J,A)=0$ and $A=2J$ into regions with different asymptotic behaviour
of $Z_2(A,J)$ at large $N$, see Fig. 1.

\begin{figure}[h!]
 \centering
   \includegraphics[width=6cm]{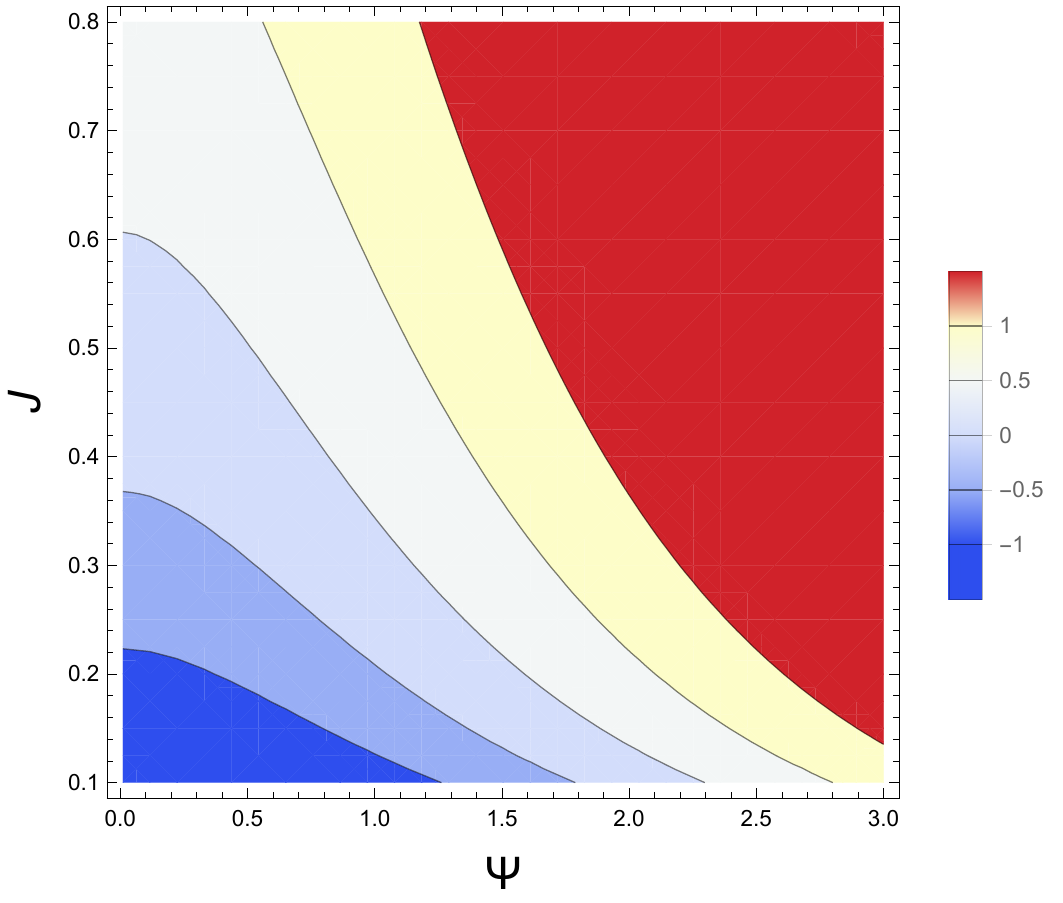}A\,\,\,\,\,\,\,
  \includegraphics[width=6cm]{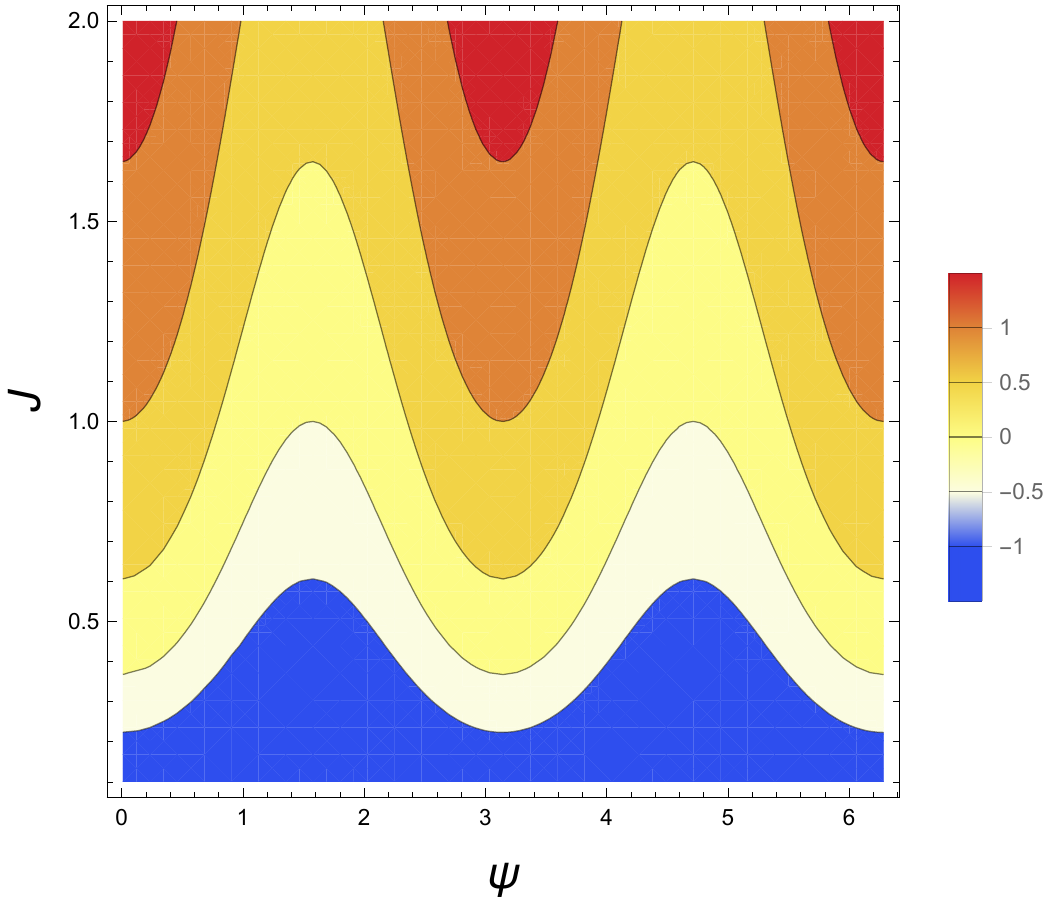}B \\$\,$\\
  \includegraphics[width=5cm]{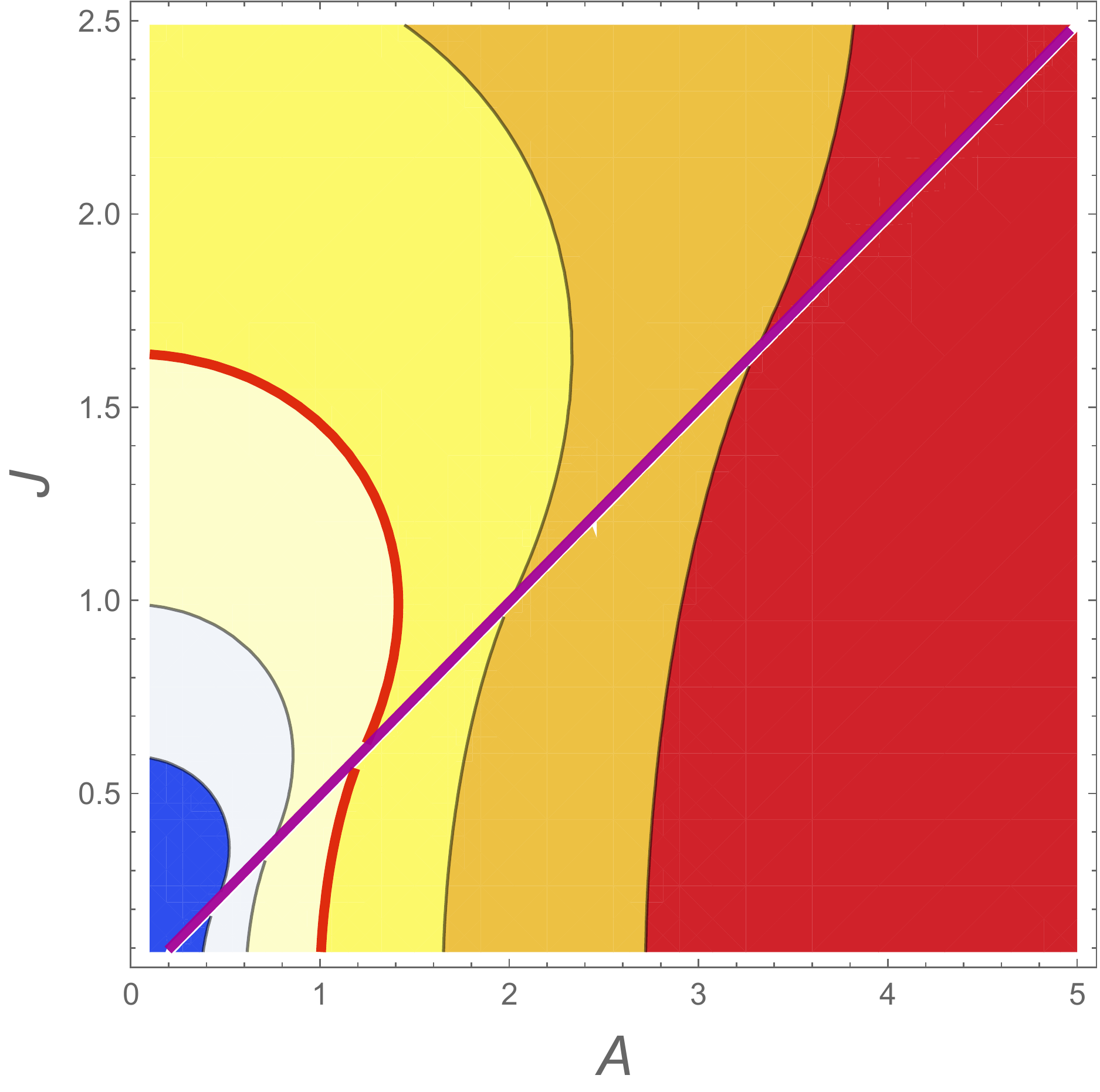}C$\,\,\,\,\,\,\,$
    \includegraphics[width=1cm]{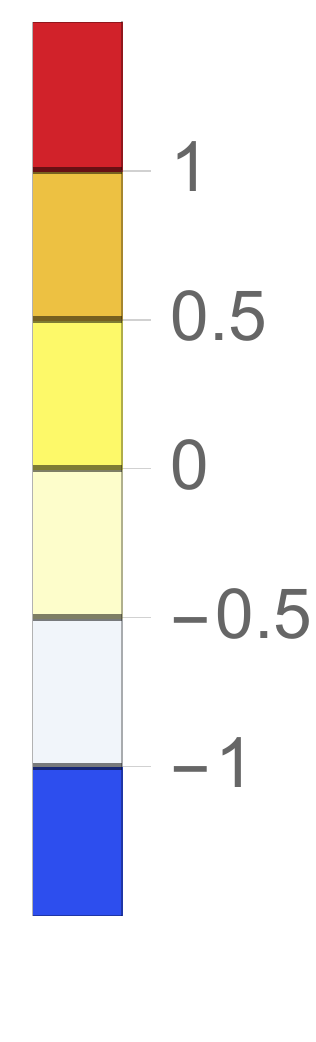}
          \caption{Contour plots for the function $F(A,J)$: A) for $A/2 J<1$ and B) $ A/2 J>1$
     as function of $\Psi,J$ and $\psi,J$, respectively. C)The contour plot for  $F(A,J)$ as function of $A$ and $J$. 
     The magenta line shows the change of regimes and the red one shows the change of sign of $F(A,J)$}
  \label{Fig:PhT1}
\end{figure}

\section{Quartic case}
\subsection{Partition function}
Let us first consider the   partition function for the quartic interaction
\be\label{F2}
Z_4(J)=\frac{N}{2\pi}\int dy\int dx\,
\exp\{N\fS_4\},\,\,\,\,\,J>0,
\ee
where
\be
\fS_4=-\frac{J^2}{4}x^4+ixy+\log y +\log i.\ee
The stationary points are derived from equations
\bea
  -J^2 x^3+i y&=&0,\\
\frac{1}{y}+i x&=&0.\eea
The solutions are
\bea
(\fx^{(1)}_0,\fy^{(1)}_0)&=&\left(-\frac{e^{\frac{i \pi }{4}}}{\sqrt{2J}}, -\sqrt{2J} e^{\frac{i \pi }{4}} \right),\\
(\fx^{(2)}_0,\fy^{(2)}_0)&=&\left(\frac{e^{\frac{i \pi }{4}}}{\sqrt{2J}}, \sqrt{2J} e^{\frac{i \pi }{4}} \right),\\
(\fx^{(3)}_0,\fy^{(3)}_0)&=&\left(- \frac{e^{\frac{i 3\pi }{4}}}{\sqrt{2J} }, \sqrt{2J} e^{\frac{i 3\pi }{4}}\right),\\
(\fx^{(4)}_0,\fy^{(4)}_0)&=&\left( \frac{e^{\frac{i 3\pi }{4}}}{\sqrt{2J} }, -\sqrt{2J} e^{\frac{i 3\pi }{4}} \right).
\eea
Expanding around the stationary points we get
\bea
&&\fS^{(1)}_{4}=\fS_{01}
-\frac32 i  J \left(x+\frac{e^{\frac{i
   \pi }{4}}}{\sqrt{2J}}\right)^2+\frac{i
   \left(y+\sqrt{2J} e^{\frac{i \pi
   }{4}} \right)^2}{2J}+i
   \left(x+\frac{e^{\frac{i \pi }{4}}}{\sqrt{2J} }\right)
   \left(y+\sqrt{2J} e^{\frac{i \pi
   }{4}} \right),\nn\\
  && \fS_{01}=\log
   \left(-\sqrt{2J}e^{\frac{i\pi }{4}}
   \right)-\frac{3}{4},\label{QF1}\eea
 \bea
&&\fS^{(2)}_{4 }=\fS_{02}-\frac32 i  J \left(x-\frac{e^{\frac{i\pi }{4}}}{\sqrt{2J}
   }\right)^2+\frac{i
   \left(y-\sqrt{2J} e^{\frac{i \pi
   }{4}} \right)^2}{2  J}+i
   \left(x-\frac{e^{\frac{i \pi
   }{4}}}{\sqrt{2J} }\right)
   \left(y-\sqrt{2J} e^{\frac{i \pi
   }{4}} \right),\nn\\&&\fS_{02}=\log
   \left( \sqrt{2J}e^{\frac{i \pi }{4}}
   \right)-\frac{3}{4},
\eea
\bea
&&\fS^{(3)}_{4}=\fS_{02}+\frac32 i J \left(x+\frac{e^{\frac{3 i
   \pi }{4}}}{\sqrt{2J}
   }\right)^2-\frac{i
   \left(y-\sqrt{2J} e^{\frac{3 i \pi
   }{4}} \right)^2}{2  J}+i
   \left(x+\frac{e^{\frac{3 i \pi
   }{4}}}{\sqrt{2J} }\right)
   \left(y-\sqrt{2J} e^{\frac{3 i \pi
   }{4}} \right),\nn\\
   &&\fS_{03}=\log
   \left( \sqrt{2J}e^{\frac{i 3\pi }{4}}
   \right)-\frac{3}{4},\eea
\bea
&&
\fS^{(4)}_{4 }=\fS_{02}+\frac32 i  J \left(x-\frac{e^{\frac{3 i
   \pi }{4}}}{\sqrt{2J}
  }\right)^2-\frac{i
   \left(y+\sqrt{2J} e^{\frac{3 i \pi
   }{4}} \right)^2}{2  J}+i
   \left(x-\frac{e^{\frac{3 i \pi
   }{4}}}{\sqrt{2J} }\right)
   \left(y+\sqrt{2J} e^{\frac{3 i \pi}{4}} \right),\nn\\
   &&\fS_{04}=\log
   \left(- \sqrt{2J}e^{\frac{i 3\pi }{4}}
  \right)-\frac{3}{4}.\eea
We note that the real parts of $\fS_{0i}, \,i=1,2,3,4$ are equal and all four points contribute to the partition function.
These contributions are the following.
 Integrating  near the first, second, third and fourth points
   \be
Z_4^{(i)}(J)=\frac{N}{2\pi}\int dy\int dx\,
\exp\{N\fS^{(i)}\},
\ee
 we get 
 \bea
 Z_4^{(1)}(J)&=&
\frac{\pi  e^{-3 N/4} \left((-1-i) \sqrt{\frac{J}{2}}\right)^N}{N},\\
Z_4^{(2)}(J)&=&
\frac{\pi  e^{-3 N/4} \left((1+i) \sqrt{\frac{J}{2}}\right)^N}{N},\\
Z_4^{(3)}(J)&=&
\frac{\pi  e^{-3 N/4} \left((-1+i)  \sqrt{\frac{J}{2}}\right)^N}{N},
\\
Z_4^{(4)}(\tilde J)&=&\frac{\pi  e^{-3 N/4} \left((1-i) \sqrt{\frac{J}{2}}\right)^N}{N}.\eea

Summing up over all critical points we obtain
\bea
Z_4(\tilde J)=\sum_{i=1}^4 Z_4^{(i)}(\tilde J)&=&  e^{-3 N/4} ( \sqrt{\frac{J}{2}})^N((-1-i)^N+(1+i)^N+(-1+i)^N+(1-i) ^N)\nn\\&=&e^{-3 N/4} J^{N/2}
\left\{
\begin{array}{ccc}
  4,& \, &   N=8n\\
 0, & \,  & N=8n+1\\
  0,& \,  &   N=8n+2\\
  0, & \,  &   N=8n+3 \\
    -4,  & ,\,  &  N=8n+4 \\
      0, & \,  &   N=8n+5 \\
          0,& \,  &  N=8n+6 \\
         0,&\,  & N=8n+7  \\
\end{array}
\right..
\eea
\subsection{Generating function}
The generating function depending on $A$ and $J$ is defined as
\be\label{F2}
Z_4(A,J)=\frac{N}{2\pi}\int dy\int dx\,
\exp\{NS_4(A,J)\},\,\,\,\,\,J>0
\ee
where
\be\label{S4JA}
S_4(A,J)=-J^2\frac{x^4}{4}+ixy+\log(i y+A).\ee
We can rewrite this expression as
\be
Z_4(A,J)=\frac{N}{2\pi}J^{N/2}\int dy\int dx\,(iy +\frac{A}{\sqrt J})^N
\exp\{ -\frac{x^4}{4}+ixy\}\ee
$$=\frac{N}{2\pi}J^{N/2}\int dy\int dx\,
\exp\{NS_4(\tilde A)
,\,\,\,\,\,J>0,
$$
where 
\be
S_4(\tilde A)=-\frac{x^4}{4}+ixy+
\log (iy+\tilde A)
\ee
and
$ \tilde A=A/\sqrt{J}.$

The stationary points are derived from equations
\bea
- x^3+i y&=&0\label{root4}\\
\frac{i}{\tilde A+i y}+i x&=&0,\eea
that give
 \bea
x^{(1)}_0&=& \frac{1}{2}M-\frac{1}{2}K_1,\,\,\,\,\,x^{(2)}_0=\frac{1}{2}M+\frac{1}{2}K_1,\nn\\
x^{(3)}_0&=&-\frac{1}{2}M-\frac{1}{2}K_2,\,\,\,\,\,\,x^{(4)}_0=-\frac{1}{2}M+\frac{1}{2}K_2,\label{x0}\eea
\bea
 y^{(1)}_0&=&  i \left(K_2
 \frac{3}{4} \tilde A+\frac{1}{8}  K_1^3+\sqrt[3]{\frac{9}{4}} \frac{
   K_1}{P}-\frac{1}{8}  M^3+ \sqrt[3]{\frac{9}{4}}\frac{
   M}{P}+\frac{1}{8} \sqrt[3]{\frac{3}{2}} K_1 P+\frac{1}{8}
   \sqrt[3]{\frac{3}{2}} M P
 \right),\nn\\ y^{(2)}_0&=& i \left(\frac{3}{4} \tilde A-\frac{1}{8}   K_1^3-\sqrt[3]{\frac{9}{4}}\frac{  K_1}{P}-\frac{1}{8}   M^3+\sqrt[3]{\frac{9}{4}}\frac{   M}{P}-\frac{1}{8} \sqrt[3]{\frac{3}{2}}
   K_1 P+\frac{1}{8} \sqrt[3]{\frac{3}{2}} M P\right),\nn\\
    y^{(3)}_0&=& i \left(\frac{3}{4} \tilde A+\frac{1}{8}   K_2^3+\sqrt[3]{\frac{9}{4}}\frac{   K_2}{P}+\frac{1}{8}  M^3-\sqrt[3]{\frac{9}{4}}\frac{   M}{P}+\frac{1}{8} \sqrt[3]{\frac{3}{2}} K_2
   P-\frac{1}{8} \sqrt[3]{\frac{3}{2}} M P\right),\nn
\\
y^{(4)}_0
&=&   i \left(\frac{3}{4} \tilde A-\frac{1}{8}   K_2^3-\sqrt[3]{\frac{9}{4}}\frac{   K_2}{P}+\frac{1}{8}  M^2-\sqrt[3]{\frac{9}{4}}\frac{   M}{P}-\frac{1}{8} \sqrt[3]{\frac{3}{2}} K_2 P-\frac{1}{8}
   \sqrt[3]{\frac{3}{2}} M P\right)\label{y0},\eea
   where
   \bea
   P&=&\sqrt[3]{\sqrt{3} \sqrt{27 \tilde A^4
   -256 }+9 \tilde A^2  },\,\,\,\,
   M=\sqrt{\frac{P}{\sqrt[3]{18}
    }+ \sqrt[3]{\frac{2}{3}}\frac{4
  }{P}},\,\,\,\,\,\\
  K_1&=&\sqrt{-\frac{2 \tilde A}{  M}-M^2},\,\,\,\,\, K_2=\sqrt{\frac{2 \tilde A}{  M}-M^2}.\eea

\begin{figure}[h!]
  \centering
   \includegraphics[width=7cm]{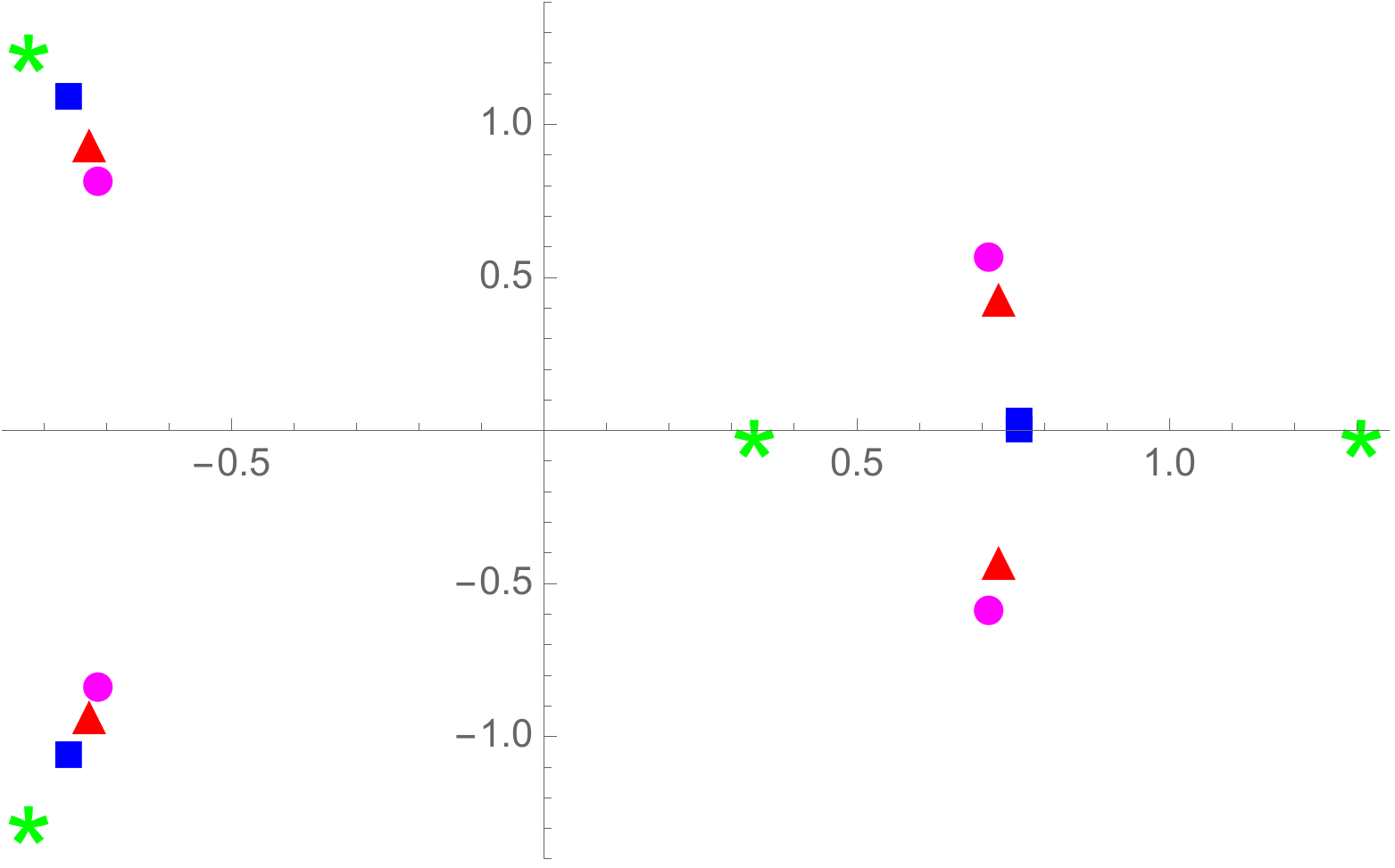}$\,\,\,\,\,\,\,\,\,\,\,\,\,$
      \includegraphics[width=7cm]{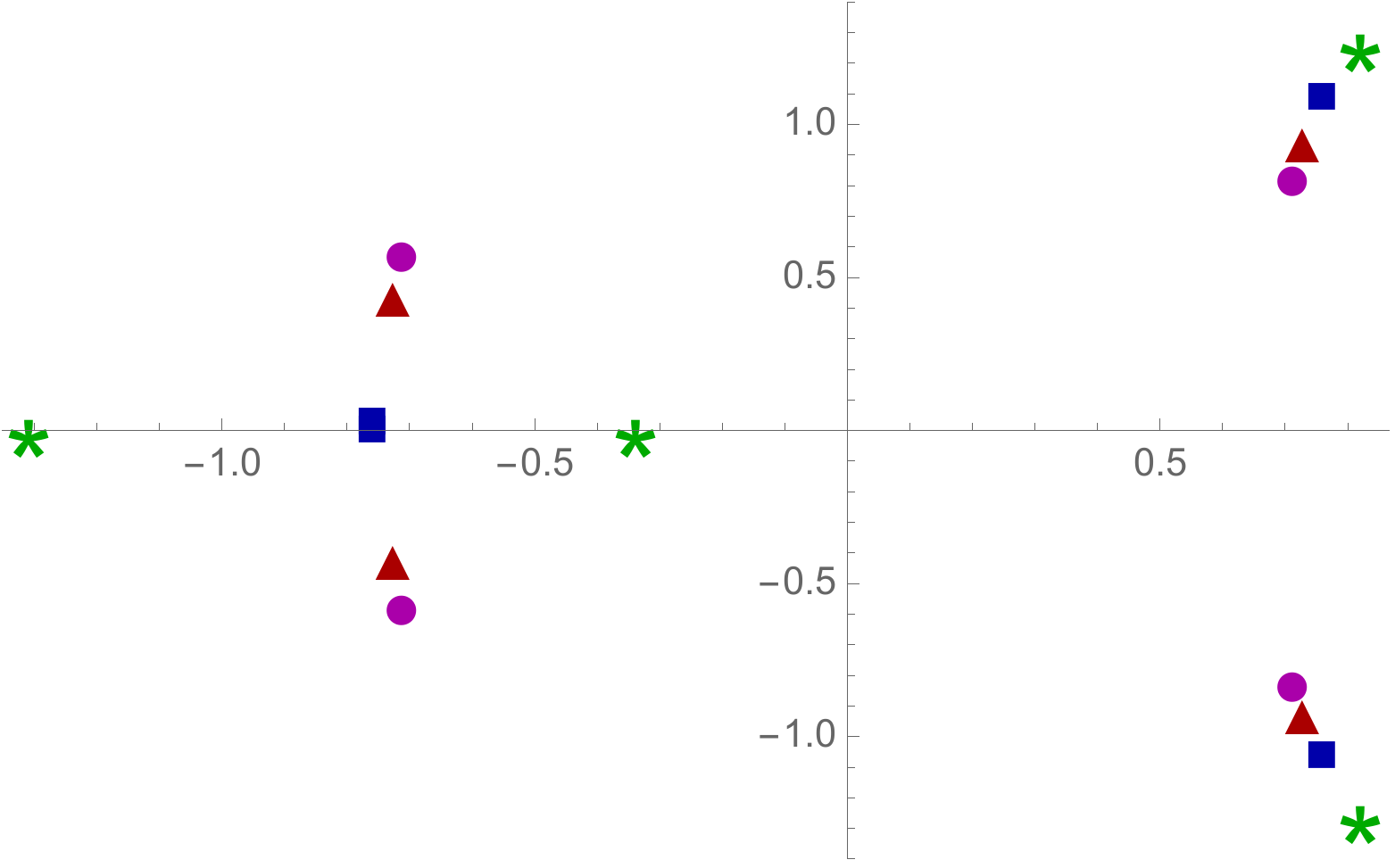}\\
          \includegraphics[width=3.5cm]{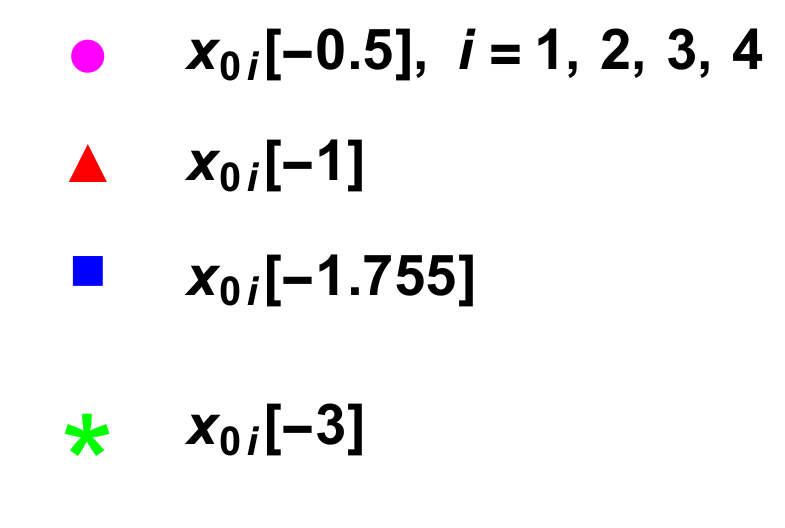}$\,\,\,\,\,\,\,\,\,\,\,\,\,$$\,\,\,\,\,\,\,\,\,\,\,\,\,$
      \includegraphics[width=3.5cm]{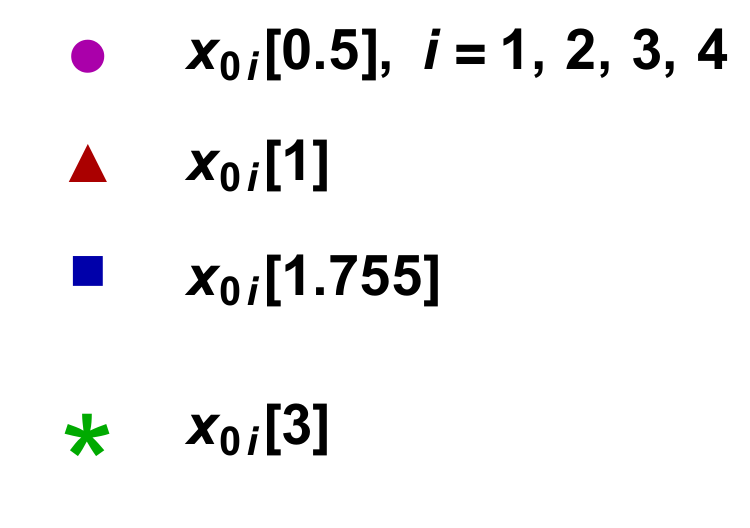}\\$$\,$$\\
         \includegraphics[width=7cm]{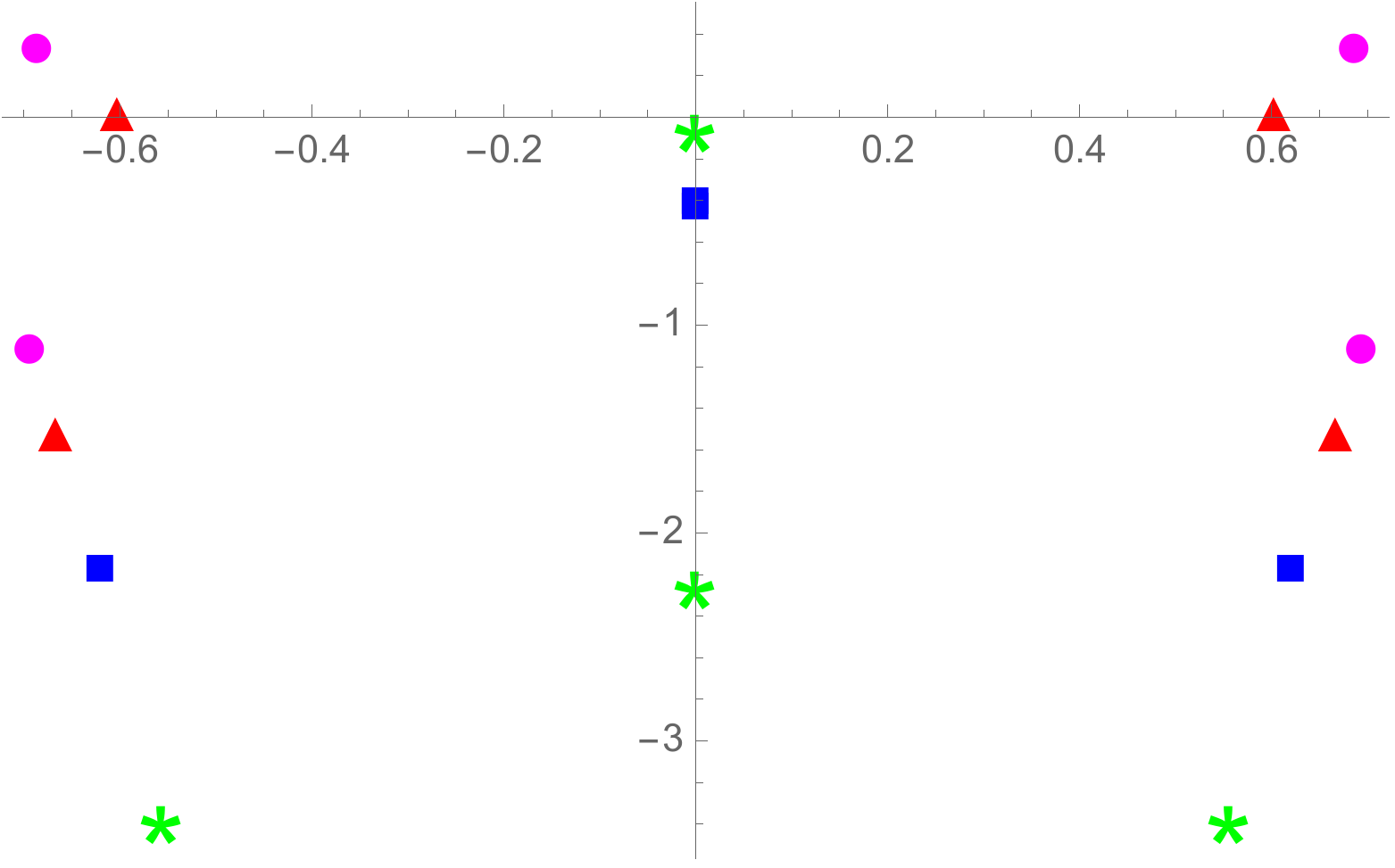}$\,\,\,\,\,\,\,\,\,\,\,\,\,$
          \includegraphics[width=7cm]{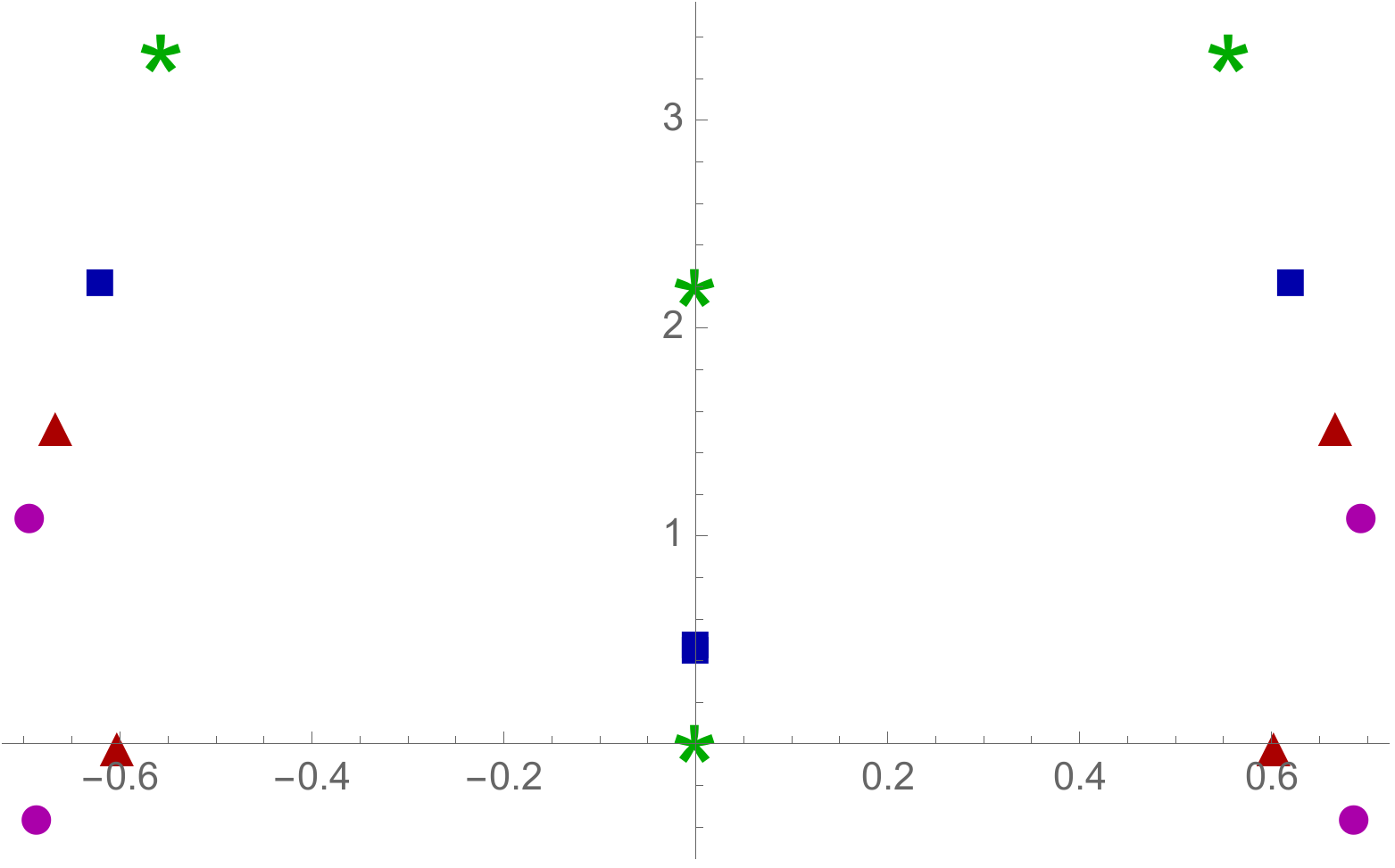}\\
      \includegraphics[width=3.5cm]{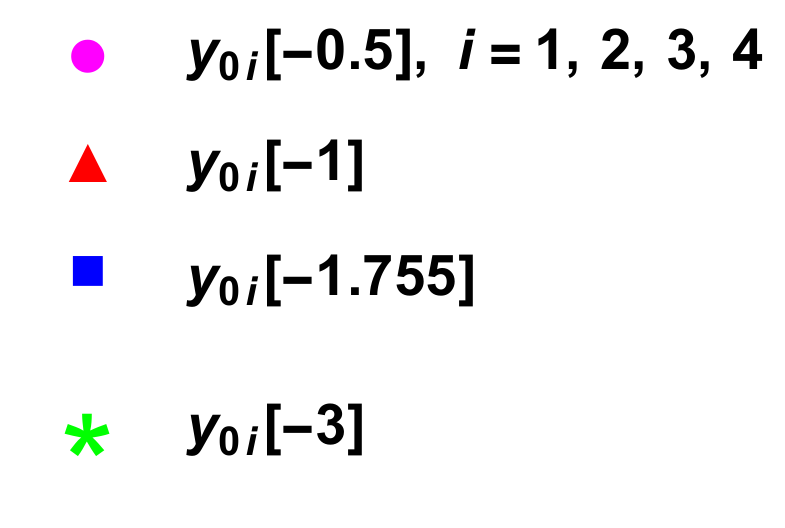}$\,\,\,\,\,\,\,\,\,\,\,\,\,$$\,\,\,\,\,\,\,\,\,\,\,\,\,$
           \includegraphics[width=3.5cm]{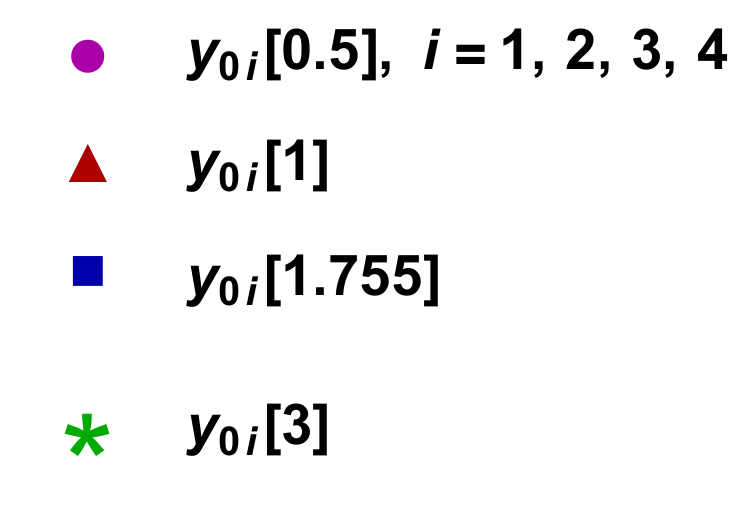}
     \caption{Location of stationary points $(x^{(i)}_0(\tilde A),\,y^{(i)}_0(\tilde A)),\,\, i=1,2,3,4$,  depending on  $\tilde A$.}
  \label{Fig:XY04}
\end{figure}

 In Fig.\ref{Fig:XY04} we present the location of these roots.
 \begin{itemize}
 \item
  We see that for $|\tilde A|<\frac{4}{3^{3/4}}$ there are two pairs of complex conjugated x-roots: $x^{(2)}_0=\bar x^{(1)}_0$,
     $x^{(4)}_0=\bar x^{(3)}_0$ and  two pairs of y-roots related as: $y^{(2)}_0=-\bar y^{(1)}_0$,
     $y^{(4)}_0=-\bar y^{(3)}_0$.  We call this domain the $\fT $ domain (the trigonometrical domain).
     \item For  $\tilde A=\pm \frac{4}{3^{3/4}}$ we have for x-roots: one pair of complex conjugated values  $x^{(1)}_0=\bar x^{(2)}_0$ and one pair  of equal reals  $x^{(3)}_0= x^{(4)}_0$, and  for y-roots:
      one pair of y-roots related as  $y^{(2)}_0=-\bar y^{(1)}_0$ and one pair of equal pure imaginary $y^{(3)}_0= y^{(4)}_0$.
      This is the critical domain $\fC $.
     \item  For $\tilde A>\frac{4}{3^{3/4}}$ we have for x-roots: one pair of complex conjugated values  $x^{(1)}_0=\bar x^{(2)}_0$ and one pair  of non-equal reals  $x^{(3)}_0< x^{(4)}_0<0$, and  for y-roots:
      one pair of y-roots related as : $y^{(2)}_0=-\bar y^{(1)}_0$ and one pair of non-equal pure imaginary $0<\Im\, y^{(4)}_0< \Im\, y^{(3)}_0$.
      This is the hyperbolic domain $\fH_+$.
      
      For $\tilde A<-\frac{4}{3^{3/4}}$ we have for x-roots: one pair of complex conjugated values  $x^{(3)}_0=\bar x^{(4)}_0$ and one pair  of non-equal reals  $0<x^{(1)}_0< x^{(2)}_0$, and  for y-roots:
      one pair of y-roots related as  $y^{(3)}_0=-\bar y^{(3)}_0$ and one pair of non-equal pure imaginary $\Im\, y^{(2)}_0< \Im\, y^{(1)}_0<0$.
       This is the hyperbolic domain $\fH_- $.

 \item      The boundaries of the $\fH_\pm$ and $\fT_\pm$  come from solutions of the equation
       \be
       K_2=0\Rightarrow 2 \tilde A-M^3=0\ee
       or explicitly
    \be
  144 \tilde A^2
   \left(\sqrt{81 \tilde A^4-768}+9 \tilde A^2\right)=\left(\sqrt[3]{2} \left(\sqrt{81 \tilde A^4-768}+9
   \tilde A^2\right)^{2/3}+8 \sqrt[3]{3}\right)^3. \label{144}\ee
   The  solutions of \eqref{144} are
   
   \be
  \tilde A_0=\pm  \frac{4}{3^{3/4}}\approx \pm 1.755.\ee
      \end{itemize}

   Expanding the action \eqref{S4JA} around these stationary points we get
   \bea
   S_4(A,J)\Big|_{x\sim x_0^{(i)}}
\approx S^{(i)}_{4}(A,J)&=&S_{0i}(A,J)+Q^{(i)}_{\alpha\beta}(\tilde A)X^{(i)\alpha }X^{(i)\beta},\,\,\,X^\alpha=(X,Y),\,\,\, i=1,2,3,4\nn\\
\label{exp}X&=&x-x_0^{(i)},\,\,\,\,Y=y-y_0^{(i)},\eea
where
\bea\label{S4JA0i}
S_{0i}(A,J)&=&S_{0i}(\tilde A)+\log \sqrt{J},\\
S_{0i}(\tilde A)&=&-\frac{x^{(i)}_0\,^4}{2}+ix^{(i)}_0y^{(i)}_0+\log(i y^{(i)}_0+\tilde A).\eea

We have two types of solutions:  trigonometric in the region  ${\fT}$ and hyperbolic in the region ${\fH}$ of the real line. In the region  ${\fT}$ there are two pairs of complex conjugated roots and
in the region ${\fH}$ there is one pair of the complex conjugated and two reals roots of \eqref{root4}.

\begin{figure}[h!]
  \centering
   \includegraphics[width=7cm]{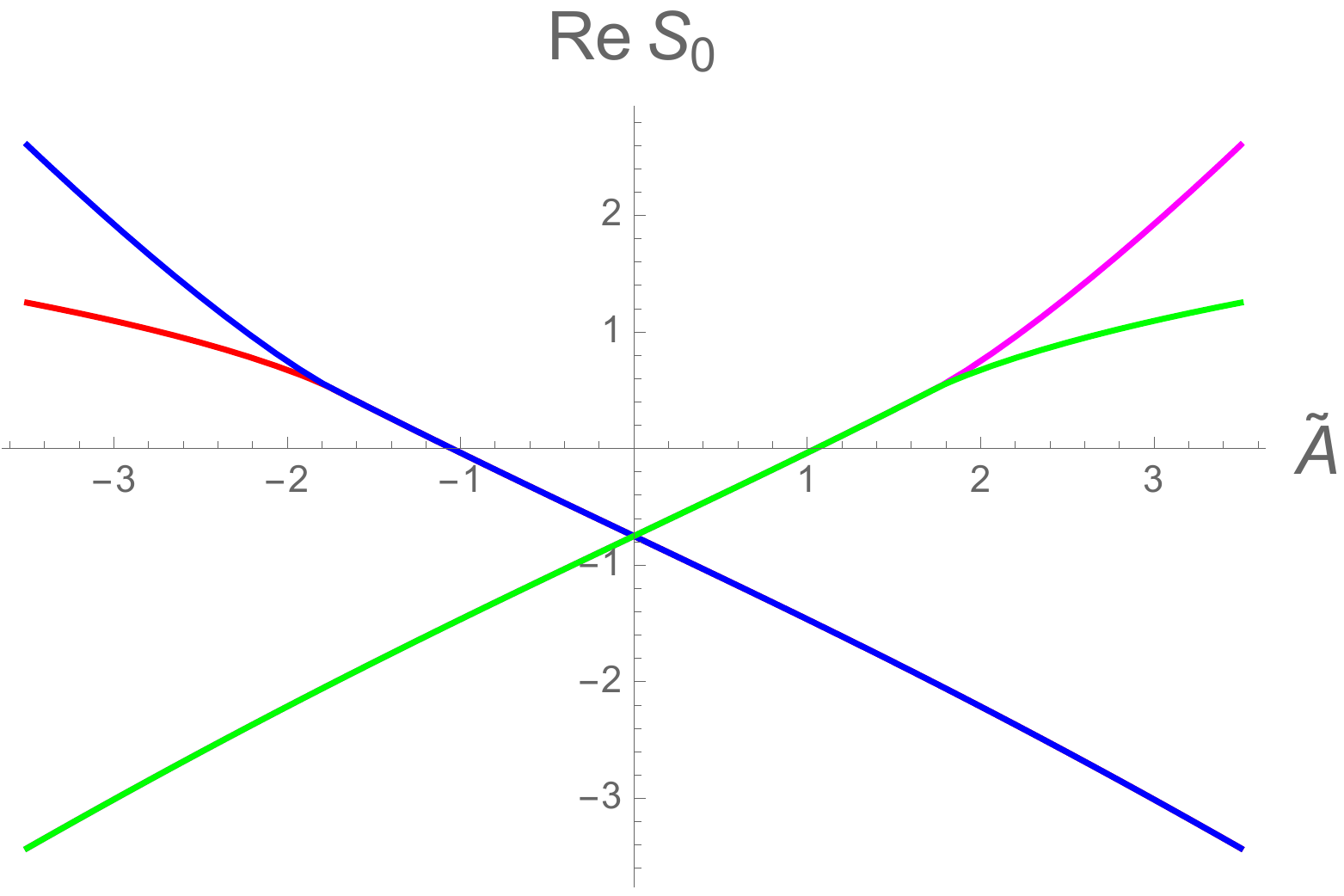}
      \includegraphics[width=2cm]{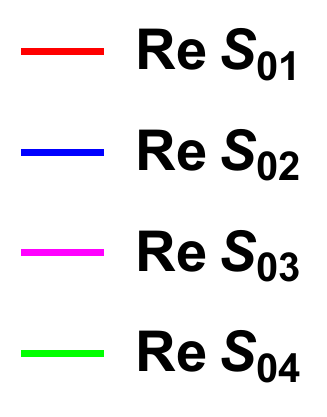}
           \caption{Values of $\Re[S_{0i}(\tilde A)], i=1,2,3,4$  for  arbitrary $\tilde A$}
  \label{Fig:ReS04}
\end{figure}

 In the domain ${\fT}$ the real parts of $S_{01}(\tilde A)$ and $S_{02}(\tilde A)$, and  the real parts of $S_{03}(\tilde A)$ and $S_{04}(\tilde A)$ coincide,
\bea
\Re[S_{01}(\tilde A)]&=&\Re[S_{02}(\tilde A)],\\
\Re[S_{03}(\tilde A)]&=&\Re[S_{04}(\tilde A)].\eea
It turns out that
\bea
\label{inH}
\Re[S_{01}(\tilde A)]&>&\Re[S_{03}(\tilde A)],\,\,\,{\mbox{ for}}\,\,\tilde A\in\fT_-,\\
\Re[S_{01}(\tilde A)]&<&\Re[S_{03}(\tilde A)],\,\,\,{\mbox{ for}}\,\, A\in\fT_+,
\eea
$\fT_\mp$ are parts of $\fT$ where $\tilde A <0$ and $\tilde A >0$, respectively.
In $\fH$ we have
\bea
\label{inT}
\Re[S_{02}(\tilde A)]&>&\Re[S_{01}(\tilde A)]>\Re[S_{03}(\tilde A)]=\Re[S_{04}(\tilde A)],\,\,\,{\mbox{ for}}\,\,\tilde A\in\fH_-,\\
\Re[S_{03}(\tilde A)]&>&\Re[S_{04}(\tilde A)]>\Re[S_{01}(\tilde A)]=\Re[S_{02}(\tilde A)],\,\,\,{\mbox{ for}}\,\,\tilde A\in\fH_+,\nn
\eea
$\fH_\mp$ are parts of $\fH$ where $\tilde A <0$ and $\tilde A >0$, respectively,
see Fig.\ref{Fig:ReS04}.

In Fig.\ref{Fig:ReS04}  we present the dependence of the $\Re[S_{0i}(\tilde A)]$, $i=1,2,3,4$, on $\tilde A$.

The quadratic forms have the following coefficients
\bea
Q^{(i)}_{11}=-\frac{3}{2}  x^{(i)}_0\,^2,\,\,\,\,Q^{(i)}_{12}=i,\,\,\,\,Q^{(i)}_{22}=\frac{1}{2 \left(\tilde A+i y^{(i)}_0\right){}^2}\eea
and integrating over $x$ and $y$ we get the prefactor
\be
\Phi ^{(k)}=\frac{2 \pi }{N \sqrt{1-\frac{3  x_0^{( k)}\,^2}{\left(\tilde A+i y_0^{(k)}\right){}^2}}},\ee
that  can be written in the form
\be
\Phi ^{(k)}=R^{(k)}e^{i\phi ^{(k)}},\,\,\,\,\,R^{(k)}>0,\,\,\,0\leq \phi ^{(k)}\leq 2\pi,\ee
$k=1,2,3,4$.

We can approximate the integral by the contributions of these  four points,
\bea\label{FFS4}
&&Z_4(A,J)=\sum_{i=1}^4\,\Phi ^{(i)}e^{NS_{0i}(A,J)},\eea
but according to the inequalities \eqref{inH} (see Fig.\ref{Fig:ReS04}) in the region $\fH_-$  the contribution from
the second point $(x^{(2)}_0,y^{(2)}_0)$ dominates, in the $\fH_+$  the contribution from
the third point $(x^{(3)}_0,y^{(3)}_0)$ dominates. In  $\fT_-$ the contributions from  the first and the second points,  $(x^{(1)}_0,y^{(1)}_0)$ and $(x^{(2)}_0,y^{(2)}_0)$  dominate, and in  $\fT_+$ 
the contributions from  the third and the forth points,  $(x^{(3)}_0,y^{(3)}_0)$ and $(x^{(4)}_0,y^{(4)}_0)$ dominate.
The final result is
\bea
Z_4(A,J)&\approx&P(\frac{ A}{\sqrt J})\,e^{NF(A,J)}\label{FA4},\eea
where
\bea
F(A,J)&\approx&
\left\{
\begin{array}{cc}
\Re S_{01}(\frac{ A}{\sqrt J})+\frac12\log J,& \,  \tilde A \in \fT_- \\
\Re S_{03}(\frac{ A}{\sqrt J})+\frac12\log J,&\,    \tilde A \in \fT_+   \\
\Re S_{01}(\frac{ A}{\sqrt J})+\frac12\log J, &       \tilde A \in \fH_-\\
\Re S_{03}(\frac{ A}{\sqrt J})+\frac12\log J, &        \tilde A \in \fH_+
\end{array}
\right.,
\label{FF4}\eea
and
\bea
P(\tilde A)=\left\{
\begin{array}{cc}
 2R^{(1)}(\tilde A)\cos(\phi^{(1)}(\tilde A)+N\Im S_{01}(\tilde A)) ,& \,  \tilde A \in \fT_- \\
  2R^{(3)}(\tilde A)\cos(\phi^{(3)}(\tilde A)+N\Im S_{03}(\tilde A)),&\,    \tilde A \in \fT_+   \\
  2R^{(1)}(\tilde A) , &       \tilde A \in \fH_-\\
  2R^{(3)}(\tilde A), &        \tilde A \in \fH_+
\end{array}
\right.
.\label{P4}\eea
\begin{figure}[h!]
  \centering
   \includegraphics[width=7cm]{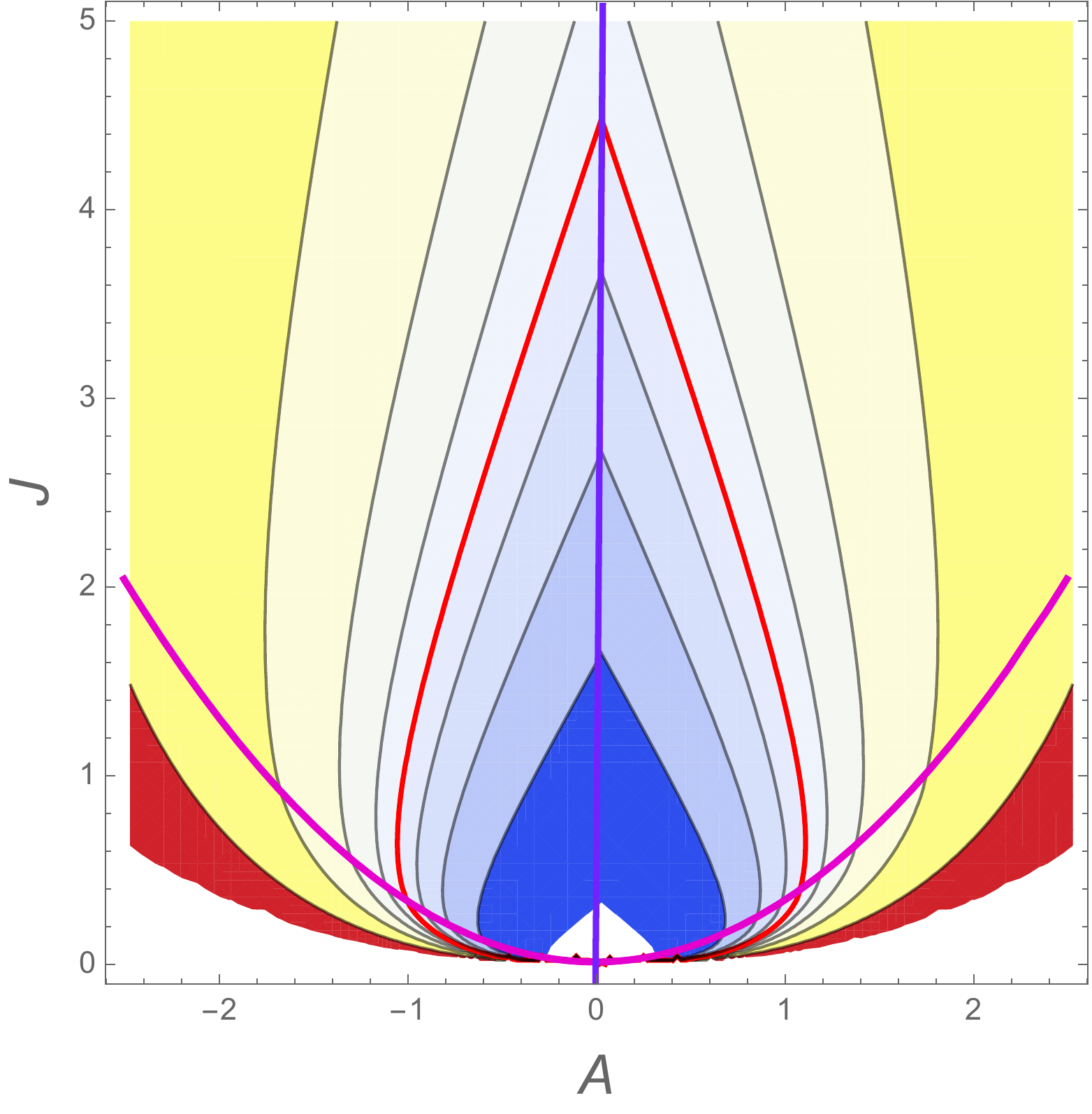}$\,\,\,\,\,\,\,$
    \includegraphics[width=1cm]{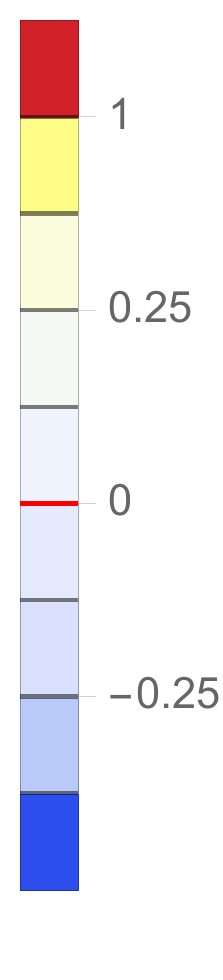}
           \caption{The equal level plot for free energy $F(A,J)$ given by \eqref{FF4}. The magenta line shows the phase transition line and the red one shows the change of sign of $F(A,J)$}
  \label{Fig:PHTR4}
\end{figure}

In Fig.\ref{Fig:PHTR4} we plot the dependence of  free energy $F(A,J)$ on parameters $A$ and $J$. We see that the half plane
$\{(A,J):\,J>0\}$ is divided  into 8 regions by critical lines.

\section{Conclusion}
The formulation of the SYK model in real time is considered. The large $N$ asymptotic behavior of the generating functional $Z_q(A,J)$ in the zero dimensional SYK model with  $M=2$ 
replicas has been studied and the  phase transitions were investigated. There are 8 regions in the upper half plane $\{(A,J):\,J>0\}$ with different behavior of the free energy $F(A,J)$. For $q=2$ the critical curves are $A^2/J^2=4,\,\, F(A,J)=0$  and for $q=4$ the critical curves are $A^4/J^2=4^4/3^{3}$ as well as  $F(A,J)=0$ and $A=0$. For the arbitrary even $q$ the 
 critical curve is given by the equation $A^q/J^2=q^q/(q-1)^{(q-1)}$. It would be interesting to study the behavior in the 
model in this transition layer, i.e. an analogue of \eqref{TRR}.
The large $N$ behavior and phase transitions in the 0-dim   SYK model in real time with arbitrary $M>2$ also deserves to study.


\section*{Acknowledgements} This work is supported by the Russian Science Foundation (project
14-50-00005, Steklov Mathematical Institute). We thank M.~Khramtsov and M.~Tikhanovskaya
for useful discussions.

 \end{document}